\documentclass[final,5p,times,twocolumn]{elsarticle}
\usepackage{amsmath}
\usepackage{tipa,tipx}

\journal{arXiv}

\begin{document}

\begin{frontmatter}

\title{Artificial neural network molecular mechanics of iron grain boundaries}

\author[1]{Yoshinori Shiihara\corref{cor1}}
\cortext[cor1]{Corresponding author}
\ead{shiihara@toyota-ti.ac.jp}
\author[1]{Ryosuke Kanazawa}
\author[2]{Daisuke Matsunaka}
\author[3]{Ivan Lobzenko}
\author[3]{Tomohito Tsuru}
\author[4]{Masanori Kohyama}
\author[5]{Hideki Mori}
\address[1]{Graduate School of Engineering, Toyota Technological Institute, Nagoya, Aichi 468-8511, Japan}
\address[2]{Department of Mechanical Systems Engineering, Shinshu University, 4-17-1 Wakasato, Nagano, Nagano 380-8553, Japan}
\address[3]{Nuclear Engineering Research Center, Japan Atomic Energy Agency, Tokai-mura, Ibaraki 319-1195, Japan}
\address[4]{Research Institute of Electrochemical Energy, Department of Energy and Environment, National Institute of Advanced Industrial Science
and Technology, Ikeda, Osaka 563-8577, Japan}
\address[5]{Department of Mechanical Engineering, College of Industrial Technology, 1-27-1 Nishikoya, Amagasaki, Hyogo 661-0047, Japan}

\begin{abstract}
This study reports grain boundary (GB) energy calculations for 46 symmetric-tilt GBs in $\alpha$-iron using molecular mechanics based on an artificial neural network (ANN) potential and compares the results with calculations based on the density functional theory (DFT), the embedded atom method (EAM), and the modified EAM (MEAM). The results by the ANN potential are in excellent agreement with those of the DFT (5\% on average), while the EAM and MEAM significantly differ from the DFT results (about 27\% on average). In a uniaxial tensile calculation of $\sum3(1\bar{1}2)$ GB, the ANN potential reproduced the brittle fracture tendency of the GB observed in the DFT while the EAM and MEAM showed mistakenly showed ductile behaviors. These results demonstrate the effectiveness of the ANN potential in grain boundary calculations of iron as a fast and accurate simulation highly in demand in the modern industrial world.
\end{abstract}

\begin{keyword}
grain boundary, artificial neural network potential, uniaxial tension, iron, molecular mechanics
\end{keyword}

\end{frontmatter}


Grain boundaries (GBs) govern the mechanical properties in polycrystalline materials, such as plasticity, material toughness, fatigue strength, creep resistance, and corrosion resistance\cite{Palumbo1998, Gleiter1989, sutton1993electronic, Lin1995, Chen2000, Moller2014a}. For example, the grain refinement can dramatically increase strength. On the other hand, brittle fracture, one of the most dangerous failures of metals, occurs at GBs. These remarkable properties stem from the rich atomic-level behavior of GBs, such as boundary sliding, migration, cleavage, solute diffusion, and responses for dislocation transmission or accumulation{\cite{Farkas2002, Langdon2006, Rupert2009, Tang2016, Sob2017, Frolov2018, Starikov2020, Tong2014}}. A GB can also be a source or sink of defects. Understanding the atomic-level behavior is essential for achieving highly functional materials through the control of GB properties.

For elucidating the dynamic behavior and structural changes of GBs at the nanoscopic level, atomic simulations are more suitable than experimental methods\cite{Sob2017,Starikov2020,Hao2014,Wang2016}. Electronic-structure based methods such as first-principles and tight-binding methods, along with molecular mechanics/dynamics based on model potentials, have been widely used to evaluate the structural stability and theoretical strength of GBs and to clarify the interaction mechanism with defects\cite{Yamaguchi2006, Bhattacharya2013,Hao2014, Bhattacharya2014, Scheiber2016, Bhattacharya2017, Wang2018, Zheng2020}. 

However, there are two complexities inherent in GBs that prevent their understanding by conventional methods. The first is their structural complexity. Even for a symmetric-tilt GB, its atomic model can consist of more than thousands of atoms depending on its misorientation angle\cite{Terentyev2010}. Moreover, since the atomic structure near the GB core is unstable, it is necessary to find the minimum value from the energy landscape with hundreds or thousands of minima in the vicinity to obtain the most stable structure\cite{Hahn2016, Ratanaphan2015}. Another complexity is its atomic-bonding state: the connectivity network in the GB core differs from the bulk. Changes in its charge transfer, magnetic state, coordination number, and misorientation angle can drastically affect the GB energy and its dynamics\cite{Bhattacharya2014, Frolov2018}. As the next step after understanding GB energies and configurations, we have to clarify the detailed procedure of dislocation transmission or accumulation at a GB, which also require much more numbers of atoms and high accuracy for various local configurations. The conventional simulations are not enough to solve these problems, as the computational time can be a critical drawback when using an electronic-structure method and model-potential methods have inherent problems with their accuracy when confronting such complex atomic systems. 

The artificial neural network (ANN) has the potential to solve these problems. This network is able to represent complicated functions as stated in the universality theorem\cite{hornik1991approximation, cybenko1989approximation}. Once constructed, the computational complexity of the ANN is proportional to the number of atoms, making it much faster compared to electronic-structure calculations. Despite the cases in fcc\cite{Nishiyama2020}, ANN and other machine learning potentials have not yet been applied to bcc GBs including GBs of $\alpha$-iron. Although iron has been one of the most important structural materials for centuries, the lack of a large-scale and accurate simulation for its GBs has hindered the solutions to many industrial problems, including hydrogen embrittlement\cite{Barrera2018}.

In this study, we perform GB calculations for $\alpha$-iron using an ANN potential designed for dislocation calculations\cite{Mori2020} and demonstrate its usefulness, especially its high accuracy and transferability, by comparing it with the density functional theory (DFT), the embedded atom method (EAM), and the modified EAM (MEAM). One of the unique features of this study is that we performed a wide range of comparisons for 46 GB systems with up to 652 atoms. This is not typical because past studies on the first-principles calculations for GBs generally treat only 200 atoms due to their computational complexity\cite{Sob2017}. In addition to the GB energy calculation, we compare surface and separation energies. We also perform uniaxial tensile simulations to compare the fracture behaviors of $\sum3(1\bar{1}2)$ GB obtained by the methods.

For the ANN potential, we used the high-dimensional neural network potential proposed by Behler, \textit{et al}\cite{Behler2007, Artrith2017}. The neural network for iron atoms was constructed by one of the authors for analyzing the motion of dislocations in Fe crystals\cite{Mori2020}. The details of the network and atomic systems considered for the training data are provided in the supplemental material, but note here that no grain-boundary data are included in the training data. The ANN package \textipa{\ae}net\cite{Artrith2016} was used for training, together with the L-BFGS optimization method\cite{byrd1995limited}. We also used LAMMPS\cite{plimpton1995fast}, a large-scale MD software, and the \textipa{\ae}net wrapper implemented in it. The \textipa{\ae}net wrapper and ANN potential files for LAMMPS used in the simulations are available online (https://github.com/HidekiMori-CIT/aenet-lammps). The potential file, Fe.10sw-10sw.nn, can be found at ANN/Fe\_v03 directory.

For comparison, calculations under the same conditions were performed using the DFT\cite{hohenberg1964, kohn1965self} and the EAM\cite{proville2012} and MEAM\cite{Asadi2015} model potentials. The first-principles results for GBs and surfaces reported below were obtained by the software package VASP\cite{kresse1993ab, kresse1999upp} based on the PAW method\cite{blochl1994projector}. The exchange-correlation functional we used is based on the Perdew-Berke-Ernzerhof generalized gradient approximation\cite{perdew1996generalized}. The cutoff energy of the plane wave was set to 400 eV, and the k-point density was set so that the grid spacing was less than $0.023 \text{ \AA}^{-1}$ with the Gaussian smoothing parameter of 0.2 eV. The convergence condition of the energy and geometrical optimization were set to be less than $10^{-6}$ eV and $0.01 \text{ eV}/\text{\AA}$ per atom, respectively. In contrast to the EAM, the MEAM can take into account the effects of oriented bonds and second nearest neighbor atoms.

The GB structures considered here are summarized in the table in the supplemental material, along with the definition of the misorientation angle. Forty-six symmetric tilt GBs ($\sum$ below 99, systems with 652 atoms at maximum) are considered. This number is far more than in previous studies\cite{Scheiber2016, Wang2018}. In the present study, which focuses on comparing methods, the atomic structure was created using a relatively simple method based on the CSL model\cite{sutton1995interfaces}. Also, for relatively unstable GBs, the energetically stable structure strongly depends on the initial structure used for the search. We set the initial structure to the stable structure obtained by the DFT calculation, where geometrical parameters (atomic positions and cell geometry along the axis perpendicular to the GB plane) are optimized. The stable structure for each method was obtained by relaxing the initial structure using the method. 

\begin{figure}[t]
\includegraphics[clip,width=8.5cm]{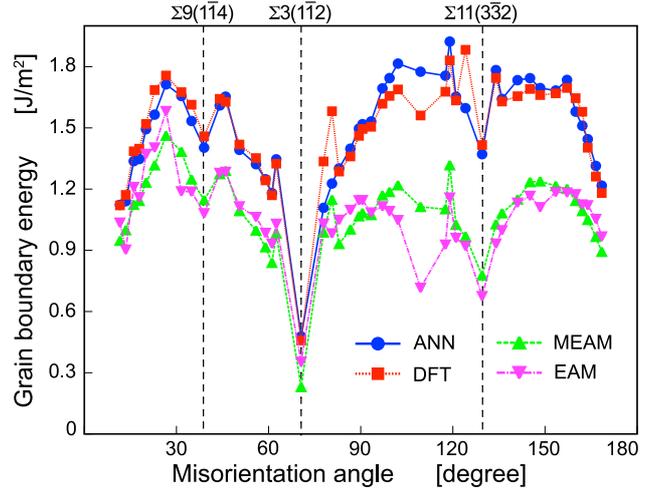}
\caption{GB energy as a function of misorientation angle calculated using DFT, ANN potential, MEAM, and EAM. The definition of the misorientation angle is given in the supplemental material.}
\label{GBE_rotangle}
\end{figure}

The surface structures used in the surface energy and separation energy calculations were created so that the boundary interfaces are exposed on the surface (see supplemental material for details). The accuracy of the surface energy is not the main focus of this study; the DFT results are given for only five points. We compare the results of tensile tests performed on $\sum3(1\bar{1}2)$ GBs by the DFT, the ANN potential, the MEAM, and the EAM. After the geometric optimization, the cell was incrementally deformed in the direction normal to the GB by 2\% until fracture occurred. For simplicity, we neglected the Poisson effect: the cell size was fixed, while the atomic structure was relaxed during the process.

Figure 1 shows the relationship between misorientation angle and GB energy. The energy curves obtained are jaggy because the searched structures are trapped in the local minima due to the simple method used to create grain-boundary models, while all the methods have similar trends. On the one hand, the GB energies of the EAM and the MEAM significantly differ from those of the DFT (about 27\% on average), as seen in the past study\cite{Scheiber2016}. In contrast, the results by the ANN potential are in excellent agreement with those of the DFT (5\% on average). To the best of our knowledge, the ANN potential in this study gives the closest GB energies of iron to those of the DFT in any atomic simulation that does not explicitly consider electrons.

Discussing the discrepancy observed in the EAM and MEAM may shed light on the origin of the GB energy. The lack of an angle-dependent term in bonding is one of the limitations of EAM; MEAM's improvement is that it takes into account the angular dependence. Since the MEAM gives different results from the DFT even with this revision, the discrepancy does not stem from this aspect. Several studies have pointed out that magnetism is the remaining problem of empirical potentials\cite{Sob2017, Ratanaphan2015}. The ANN potential is in excellent agreement with the DFT, possibly because it includes data on lattice defects such as the surfaces, vacancies, and unstable atomic structures in the gamma surface calculations as its training data, where the change in magnetism due to defects is implicitly incorporated. The TB model, which incorporates spin, reproduces the GB energy of iron\cite{Wang2018}. However, we emphasize that the TB model includes the GB energy in the fitting data, while the ANN potential does not. These findings demonstrate the excellent transferability of the ANN potential.

Figure 2 shows the relationship between the misorientation angle and the surface energy. We can see here that the ANN results are again consistent with those of the DFT, while EAM does not pass the test. Unlike the GB energy case, the MEAM gives results close to the DFT and ANN results. Since the MEAM solves the surface energy problem in the EAM, the surface energy is largely affected by the angular dependence of the bonding and/or the effect from the second neighbors in a bcc crystal. The separation energy can be evaluated from the GB and surface energies (Fig. 3). The contribution from the surface energy is more dominant than that from the GB energy. Hence, the error of the MEAM in the GB energy is less problematic in the case of the separation energy calculation.

\begin{figure}[h]
\includegraphics[clip,width=8.5cm]{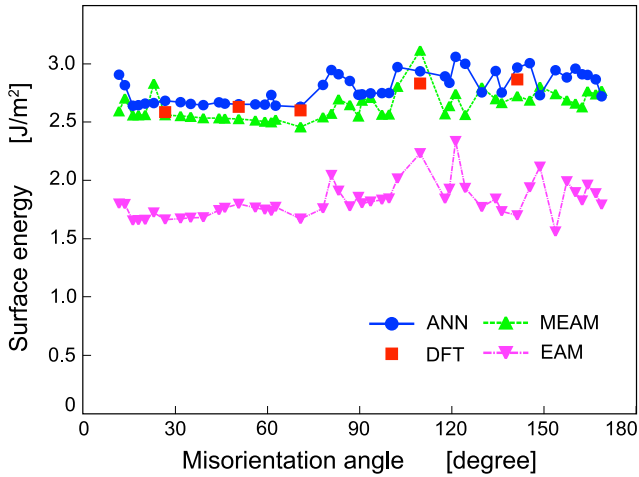}
\caption{Surface energy as a function of misorientation angle calculated using DFT, ANN potential, MEAM, and EAM.}
\label{SurE_rotangle}
\end{figure}

\begin{figure}[h]
\includegraphics[clip,width=8.5cm]{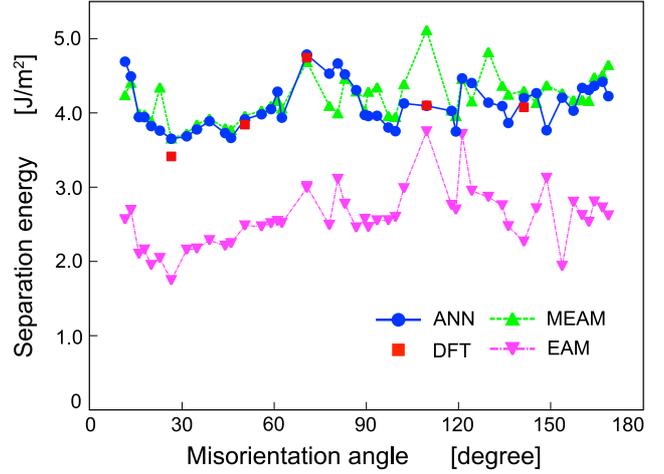}
\caption{Separation energy as a function of misorientation angle calculated using DFT, ANN potential, MEAM, and EAM.}
\label{SepaE_rotangle}
\end{figure}

\begin{figure}[h]
\includegraphics[clip,width=8.5cm]{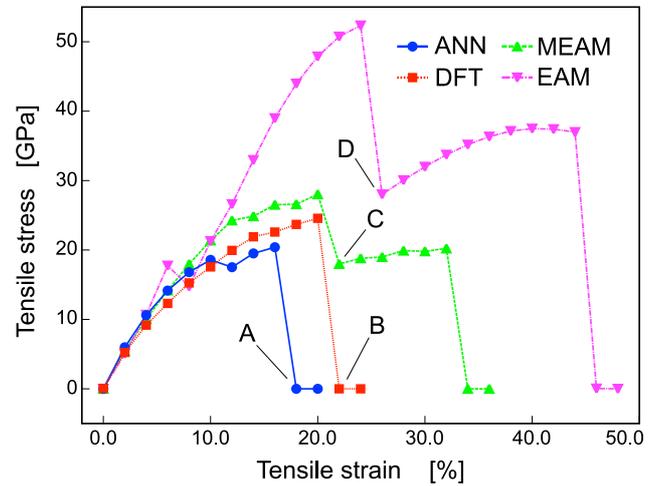}
\caption{Comparison of stress-strain curves in Fe GB $\sum3(1\bar{1}2)$ under uniaxial strain. Events A--D correspond to snapshots (a)--(d) in Fig. 5.}
\label{Stress_strain}
\end{figure}

\begin{figure}[h]
\includegraphics[clip,width=8.5cm]{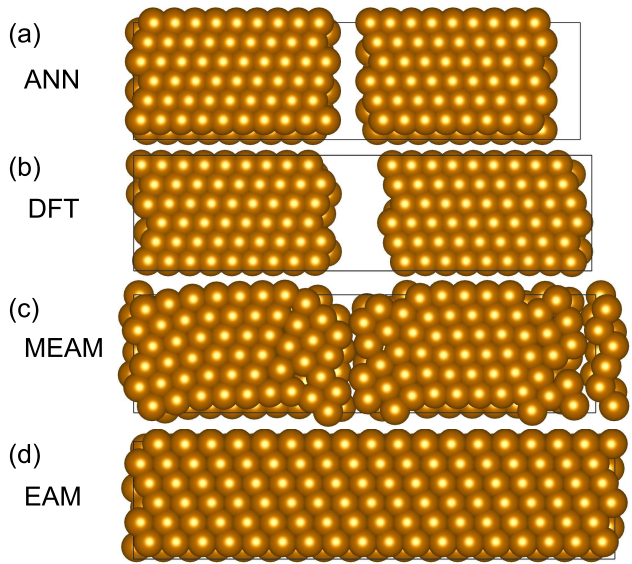}
\caption{Comparison of fracture behaviors in Fe GB $\sum3(1\bar{1}2)$ under uniaxial strain. Snapshots (a)--(d) correspond to events A--D in Fig. 4. The atomic structure is visualized by VESTA\cite{momma2011vesta}.}
\label{Atomic_structure_tension}
\end{figure}

The MEAM seems accurate enough with separation energy but not with the expression of fracture behavior. Figure 4 compares the uniaxial tensile test results of the different methods. As in the previous study\cite{yuasa2010a}, the DFT showed a brittle fracture (Fig. \ref{Atomic_structure_tension}(b)), while the MEAM and EAM showed an extensive reconstruction at intergranular fracture (Fig. \ref{Atomic_structure_tension}(c) and (d)). This reconstruction leads to an excessive maximum fracture strain, resulting in a significant difference in the stress-strain curves. The GB in the EAM disappears after several sets of phase transformation because the potential is angular independent. This is not the case with MEAM, which shows large reconstruction under uniaxial deformation, inducing the ductile behavior. The discrepancy with the separation energy comes from the fact that the energy calculation does not consider reconstruction\cite{Terentyev2010}. These ductile behaviors of the model potentials are also observed in crack growth in bcc iron crystals and are considered problematic\cite{Moller2014a, Suzudo2020}. The ANN potential underestimates the maximum fracture strain by about 4\% strain compared to the DFT but shows a similar brittle fracture (Fig. \ref{Atomic_structure_tension}(a)). Although additional potential training is needed to improve the accuracy, the results show that the ANN potential, involving the training datasets for lattice defects, keeps its transferability even for strongly nonlinear phenomena such as fracture.

In this work, we calculated GB energies for 46 cases of symmetric tilt GBs of $\alpha$-iron using the ANN potential and compared the results with DFT, MEAM, and EAM. The ANN potential reproduced the DFT results at an unachievable level with existing model potentials, even though the training data do not include the GB energies themselves. For surface energy and separation energy of the GBs, the ANN potential also showed excellent accuracy. Furthermore, in the uniaxial tensile test on the $\sum3(1\bar{1}2)$ GB, the ANN potential predicted brittle fracture similar to the DFT, whereas other model potentials mistakenly predicted ductile fracture. We believe that the ANN potential, whose accuracy at GBs has been demonstrated in this study, is a powerful tool to tackle challenging problems in iron, such as hydrogen embrittlement.

\section*{Acknowledgement}
One of the authors (H.M.) would like to thank Drs. Okumura and Itakura at JAEA for their advises about the potential development. This study was partially supported by JSPS KAKENHI Grant Numbers JP 21H00108 and 21H00154.


\begin{thebibliography}{48}
\expandafter\ifx\csname natexlab\endcsname\relax\def\natexlab#1{#1}\fi
\providecommand{\url}[1]{\texttt{#1}}
\providecommand{\href}[2]{#2}
\providecommand{\path}[1]{#1}
\providecommand{\DOIprefix}{doi:}
\providecommand{\ArXivprefix}{arXiv:}
\providecommand{\URLprefix}{URL: }
\providecommand{\Pubmedprefix}{pmid:}
\providecommand{\doi}[1]{\href{http://dx.doi.org/#1}{\path{#1}}}
\providecommand{\Pubmed}[1]{\href{pmid:#1}{\path{#1}}}
\providecommand{\bibinfo}[2]{#2}
\ifx\xfnm\relax \def\xfnm[#1]{\unskip,\space#1}\fi
\bibitem[{Palumbo et~al.(1998)Palumbo, Lehockey, and Lin}]{Palumbo1998}
\bibinfo{author}{G.~Palumbo}, \bibinfo{author}{E.~M. Lehockey},
  \bibinfo{author}{P.~Lin},
\newblock \bibinfo{journal}{JOM} \bibinfo{volume}{50} (\bibinfo{year}{1998})
  \bibinfo{pages}{40--43}.
\bibitem[{Gleiter(1989)}]{Gleiter1989}
\bibinfo{author}{H.~Gleiter},
\newblock \bibinfo{journal}{Progress in Materials Science} \bibinfo{volume}{33}
  (\bibinfo{year}{1989}) \bibinfo{pages}{223--315}.
\bibitem[{Sutton(1993)}]{sutton1993electronic}
\bibinfo{author}{A.~P. Sutton}, \bibinfo{title}{{Electronic structure of
  materials}}, \bibinfo{publisher}{Oxford University Press, USA},
  \bibinfo{year}{1993}.
\bibitem[{Lin et~al.(1995)Lin, Palumbo, Erb, and Aust}]{Lin1995}
\bibinfo{author}{P.~Lin}, \bibinfo{author}{G.~Palumbo},
  \bibinfo{author}{U.~Erb}, \bibinfo{author}{K.~T. Aust},
\newblock \bibinfo{journal}{Scripta Metallurgica et Materiala}
  \bibinfo{volume}{33} (\bibinfo{year}{1995}) \bibinfo{pages}{1387--1392}.
\bibitem[{Chen et~al.(2000)Chen, Sixta, Zhang, {De Jonghe}, and
  Ritchie}]{Chen2000}
\bibinfo{author}{D.~Chen}, \bibinfo{author}{M.~E. Sixta},
  \bibinfo{author}{X.~F. Zhang}, \bibinfo{author}{L.~C. {De Jonghe}},
  \bibinfo{author}{R.~O. Ritchie},
\newblock \bibinfo{journal}{Acta Materialia} \bibinfo{volume}{48}
  (\bibinfo{year}{2000}) \bibinfo{pages}{4599--4608}.
\bibitem[{M{\"{o}}ller and Bitzek(2014)}]{Moller2014a}
\bibinfo{author}{J.~J. M{\"{o}}ller}, \bibinfo{author}{E.~Bitzek},
\newblock \bibinfo{journal}{Acta Materialia} \bibinfo{volume}{73}
  (\bibinfo{year}{2014}) \bibinfo{pages}{1--11}.
\bibitem[{Farkas et~al.(2002)Farkas, {Van Swygenhoven}, and
  Derlet}]{Farkas2002}
\bibinfo{author}{D.~Farkas}, \bibinfo{author}{H.~{Van Swygenhoven}},
  \bibinfo{author}{P.~M. Derlet},
\newblock \bibinfo{journal}{Physical Review B} \bibinfo{volume}{66}
  (\bibinfo{year}{2002}) \bibinfo{pages}{601011--601014}.
\bibitem[{Langdon(2006)}]{Langdon2006}
\bibinfo{author}{T.~G. Langdon},
\newblock \bibinfo{journal}{Journal of Materials Science} \bibinfo{volume}{41}
  (\bibinfo{year}{2006}) \bibinfo{pages}{597--609}.
\bibitem[{Rupert et~al.(2009)Rupert, Gianola, Gan, and Hemker}]{Rupert2009}
\bibinfo{author}{T.~J. Rupert}, \bibinfo{author}{D.~S. Gianola},
  \bibinfo{author}{Y.~Gan}, \bibinfo{author}{K.~J. Hemker},
\newblock \bibinfo{journal}{Science} \bibinfo{volume}{326}
  (\bibinfo{year}{2009}) \bibinfo{pages}{1686--1690}.
\bibitem[{Tang et~al.(2016)Tang, Guo, Fan, Yip, and Yildiz}]{Tang2016}
\bibinfo{author}{X.-Z. Tang}, \bibinfo{author}{Y.-F. Guo},
  \bibinfo{author}{Y.~Fan}, \bibinfo{author}{S.~Yip},
  \bibinfo{author}{B.~Yildiz},
\newblock \bibinfo{journal}{Acta Materialia} \bibinfo{volume}{105}
  (\bibinfo{year}{2016}) \bibinfo{pages}{147--154}.
\bibitem[{Lej{\v{c}}ek et~al.(2017)Lej{\v{c}}ek, Paidar, and
  {\v{S}}ob}]{Sob2017}
\bibinfo{author}{P.~Lej{\v{c}}ek}, \bibinfo{author}{V.~Paidar},
  \bibinfo{author}{M.~{\v{S}}ob},
\newblock \bibinfo{journal}{Progress in Materials Science} \bibinfo{volume}{87}
  (\bibinfo{year}{2017}) \bibinfo{pages}{83--139}.
\bibitem[{Frolov et~al.(2018)Frolov, Rudd, Zhu, and Marian}]{Frolov2018}
\bibinfo{author}{T.~Frolov}, \bibinfo{author}{R.~E. Rudd},
  \bibinfo{author}{Q.~Zhu}, \bibinfo{author}{J.~Marian},
\newblock \bibinfo{journal}{Nanoscale} \bibinfo{volume}{10}
  (\bibinfo{year}{2018}) \bibinfo{pages}{8253--8268}.
\bibitem[{Starikov et~al.(2020)Starikov, Mrovec, and Drautz}]{Starikov2020}
\bibinfo{author}{S.~Starikov}, \bibinfo{author}{M.~Mrovec},
  \bibinfo{author}{R.~Drautz},
\newblock \bibinfo{journal}{Acta Materialia} \bibinfo{volume}{188}
  (\bibinfo{year}{2020}) \bibinfo{pages}{560--569}.
\bibitem[{Tong et~al.(2014)Tong, Zhang, and Li}]{Tong2014}
\bibinfo{author}{X.~Tong}, \bibinfo{author}{H.~Zhang}, \bibinfo{author}{D.~Li},
\newblock \bibinfo{journal}{Modelling and Simulation in Materials Science and
  Engineering} \bibinfo{volume}{22} (\bibinfo{year}{2014})
  \bibinfo{pages}{1--15}.
\bibitem[{Hao et~al.(2014)Hao, Elfimov, and Militzer}]{Hao2014}
\bibinfo{author}{J.~Hao}, \bibinfo{author}{I.~Elfimov},
  \bibinfo{author}{M.~Militzer},
\newblock \bibinfo{journal}{Journal of Applied Physics} \bibinfo{volume}{5}
  (\bibinfo{year}{2014}) \bibinfo{pages}{093506}.
\bibitem[{Wang et~al.(2016)Wang, Janisch, Madsen, and Drautz}]{Wang2016}
\bibinfo{author}{J.~Wang}, \bibinfo{author}{R.~Janisch}, \bibinfo{author}{G.~K.
  Madsen}, \bibinfo{author}{R.~Drautz},
\newblock \bibinfo{journal}{Acta Materialia} \bibinfo{volume}{115}
  (\bibinfo{year}{2016}) \bibinfo{pages}{259--268}.
\bibitem[{Yamaguchi et~al.(2006)Yamaguchi, Shiga, and Kaburaki}]{Yamaguchi2006}
\bibinfo{author}{M.~Yamaguchi}, \bibinfo{author}{M.~Shiga},
  \bibinfo{author}{H.~Kaburaki},
\newblock \bibinfo{journal}{Materials Transactions} \bibinfo{volume}{47}
  (\bibinfo{year}{2006}) \bibinfo{pages}{2682--2689}.
\bibitem[{Bhattacharya et~al.(2013)Bhattacharya, Tanaka, Shiihara, and
  Kohyama}]{Bhattacharya2013}
\bibinfo{author}{S.~K. Bhattacharya}, \bibinfo{author}{S.~Tanaka},
  \bibinfo{author}{Y.~Shiihara}, \bibinfo{author}{M.~Kohyama},
\newblock \bibinfo{journal}{Journal of Physics: Condensed Matter}
  \bibinfo{volume}{25} (\bibinfo{year}{2013}) \bibinfo{pages}{135004}.
\bibitem[{Bhattacharya et~al.(2014)Bhattacharya, Tanaka, Shiihara, and
  Kohyama}]{Bhattacharya2014}
\bibinfo{author}{S.~K. Bhattacharya}, \bibinfo{author}{S.~Tanaka},
  \bibinfo{author}{Y.~Shiihara}, \bibinfo{author}{M.~Kohyama},
\newblock \bibinfo{journal}{Journal of Materials Science} \bibinfo{volume}{49}
  (\bibinfo{year}{2014}) \bibinfo{pages}{3980--3995}.
\bibitem[{Scheiber et~al.(2016)Scheiber, Pippan, Puschnig, and
  Romaner}]{Scheiber2016}
\bibinfo{author}{D.~Scheiber}, \bibinfo{author}{R.~Pippan},
  \bibinfo{author}{P.~Puschnig}, \bibinfo{author}{L.~Romaner},
\newblock \bibinfo{journal}{Modelling and Simulation in Materials Science and
  Engineering} \bibinfo{volume}{24} (\bibinfo{year}{2016})
  \bibinfo{pages}{035013}.
\bibitem[{Bhattacharya et~al.(2017)Bhattacharya, Kohyama, Tanaka, and
  Shiihara}]{Bhattacharya2017}
\bibinfo{author}{S.~K. Bhattacharya}, \bibinfo{author}{M.~Kohyama},
  \bibinfo{author}{S.~Tanaka}, \bibinfo{author}{Y.~Shiihara},
\newblock \bibinfo{journal}{Materials Research Express} \bibinfo{volume}{4}
  (\bibinfo{year}{2017}) \bibinfo{pages}{116518}.
\bibitem[{Wang et~al.(2018)Wang, Madsen, and Drautz}]{Wang2018}
\bibinfo{author}{J.~Wang}, \bibinfo{author}{G.~K.~H. Madsen},
  \bibinfo{author}{R.~Drautz},
\newblock \bibinfo{journal}{Modelling and Simulation in Materials Science and
  Engineering} \bibinfo{volume}{26} (\bibinfo{year}{2018})
  \bibinfo{pages}{25008}.
\bibitem[{Zheng et~al.(2020)Zheng, Li, Tran, Chen, Horton, Winston, Aslaug, and
  Ping}]{Zheng2020}
\bibinfo{author}{H.~Zheng}, \bibinfo{author}{X.-G. Li},
  \bibinfo{author}{R.~Tran}, \bibinfo{author}{C.~Chen},
  \bibinfo{author}{M.~Horton}, \bibinfo{author}{D.~Winston},
  \bibinfo{author}{K.~Aslaug}, \bibinfo{author}{S.~Ping},
\newblock \bibinfo{journal}{Acta Materialia} \bibinfo{volume}{186}
  (\bibinfo{year}{2020}) \bibinfo{pages}{40--49}.
\bibitem[{Terentyev et~al.(2010)Terentyev, He, Serra, and
  Kuriplach}]{Terentyev2010}
\bibinfo{author}{D.~Terentyev}, \bibinfo{author}{X.~He},
  \bibinfo{author}{A.~Serra}, \bibinfo{author}{J.~Kuriplach},
\newblock \bibinfo{journal}{Computational Materials Science}
  \bibinfo{volume}{49} (\bibinfo{year}{2010}) \bibinfo{pages}{419--429}.
\bibitem[{Hahn et~al.(2016)Hahn, Fensin, Germann, and Meyers}]{Hahn2016}
\bibinfo{author}{E.~N. Hahn}, \bibinfo{author}{S.~J. Fensin},
  \bibinfo{author}{T.~C. Germann}, \bibinfo{author}{M.~A. Meyers},
\newblock \bibinfo{journal}{Scripta Materialia} \bibinfo{volume}{116}
  (\bibinfo{year}{2016}) \bibinfo{pages}{108--111}.
\bibitem[{Ratanaphan et~al.(2015)Ratanaphan, Olmsted, Bulatov, Holm, Rollett,
  and Rohrer}]{Ratanaphan2015}
\bibinfo{author}{S.~Ratanaphan}, \bibinfo{author}{D.~L. Olmsted},
  \bibinfo{author}{V.~V. Bulatov}, \bibinfo{author}{E.~A. Holm},
  \bibinfo{author}{A.~D. Rollett}, \bibinfo{author}{G.~S. Rohrer},
\newblock \bibinfo{journal}{Acta Materialia} \bibinfo{volume}{88}
  (\bibinfo{year}{2015}) \bibinfo{pages}{346--354}.
\bibitem[{Hornik(1991)}]{hornik1991approximation}
\bibinfo{author}{K.~Hornik},
\newblock \bibinfo{journal}{Neural Networks} \bibinfo{volume}{4}
  (\bibinfo{year}{1991}) \bibinfo{pages}{251--257}.
\bibitem[{Cybenko(1989)}]{cybenko1989approximation}
\bibinfo{author}{G.~Cybenko},
\newblock \bibinfo{journal}{Mathematics of Control, Signals and Systems}
  \bibinfo{volume}{2} (\bibinfo{year}{1989}) \bibinfo{pages}{303--314}.
\bibitem[{Nishiyama et~al.(2020)Nishiyama, Seko, and Tanaka}]{Nishiyama2020}
\bibinfo{author}{T.~Nishiyama}, \bibinfo{author}{A.~Seko},
  \bibinfo{author}{I.~Tanaka},
\newblock \bibinfo{journal}{Physical Review Materials} \bibinfo{volume}{4}
  (\bibinfo{year}{2020}) \bibinfo{pages}{123607}.
\bibitem[{Barrera et~al.(2018)Barrera, Bombac, Chen, Daff, Gong, and
  Haley}]{Barrera2018}
\bibinfo{author}{O.~Barrera}, \bibinfo{author}{D.~Bombac},
  \bibinfo{author}{Y.~Chen}, \bibinfo{author}{T.~D. Daff},
  \bibinfo{author}{P.~Gong}, \bibinfo{author}{D.~Haley},
\newblock \bibinfo{journal}{Journal of Materials Science} \bibinfo{volume}{53}
  (\bibinfo{year}{2018}) \bibinfo{pages}{6251--6290}.
\bibitem[{Mori and Ozaki(2020)}]{Mori2020}
\bibinfo{author}{H.~Mori}, \bibinfo{author}{T.~Ozaki},
\newblock \bibinfo{journal}{Physical Review Materials} \bibinfo{volume}{4}
  (\bibinfo{year}{2020}) \bibinfo{pages}{40601}.
\bibitem[{Behler and Parrinello(2007)}]{Behler2007}
\bibinfo{author}{J.~Behler}, \bibinfo{author}{M.~Parrinello},
\newblock \bibinfo{journal}{Physical Review Letters} \bibinfo{volume}{98}
  (\bibinfo{year}{2007}) \bibinfo{pages}{1--4}.
\bibitem[{Artrith et~al.(2017)Artrith, Urban, and Ceder}]{Artrith2017}
\bibinfo{author}{N.~Artrith}, \bibinfo{author}{A.~Urban},
  \bibinfo{author}{G.~Ceder},
\newblock \bibinfo{journal}{Physical Review B} \bibinfo{volume}{014112}
  (\bibinfo{year}{2017}) \bibinfo{pages}{1--5}.
\bibitem[{Artrith and Urban(2016)}]{Artrith2016}
\bibinfo{author}{N.~Artrith}, \bibinfo{author}{A.~Urban},
\newblock \bibinfo{journal}{Computational Materials Science}
  \bibinfo{volume}{114} (\bibinfo{year}{2016}) \bibinfo{pages}{135--150}.
\bibitem[{Byrd et~al.(1995)Byrd, Lu, Nocedal, and Zhu}]{byrd1995limited}
\bibinfo{author}{R.~H. Byrd}, \bibinfo{author}{P.~Lu},
  \bibinfo{author}{J.~Nocedal}, \bibinfo{author}{C.~Zhu},
\newblock \bibinfo{journal}{SIAM Journal on Scientific Computing}
  \bibinfo{volume}{16} (\bibinfo{year}{1995}) \bibinfo{pages}{1190--1208}.
\bibitem[{Plimpton(1995)}]{plimpton1995fast}
\bibinfo{author}{S.~Plimpton},
\newblock \bibinfo{journal}{Journal of Computational Physics}
  \bibinfo{volume}{117} (\bibinfo{year}{1995}) \bibinfo{pages}{1--19}.
\bibitem[{Hohenberg and Kohn(1964)}]{hohenberg1964}
\bibinfo{author}{P.~Hohenberg}, \bibinfo{author}{W.~Kohn},
\newblock \bibinfo{journal}{Physical Review} \bibinfo{volume}{136}
  (\bibinfo{year}{1964}) \bibinfo{pages}{B864}.
\bibitem[{Kohn and Sham(1965)}]{kohn1965self}
\bibinfo{author}{W.~Kohn}, \bibinfo{author}{L.~J. Sham},
\newblock \bibinfo{journal}{Physical Review} \bibinfo{volume}{140}
  (\bibinfo{year}{1965}) \bibinfo{pages}{A1133}.
\bibitem[{Proville et~al.(2012)Proville, Rodney, and Marinica}]{proville2012}
\bibinfo{author}{L.~Proville}, \bibinfo{author}{D.~Rodney},
  \bibinfo{author}{M.-c. Marinica},
\newblock \bibinfo{journal}{Nature Materials} \bibinfo{volume}{11}
  (\bibinfo{year}{2012}) \bibinfo{pages}{845--849}.
\bibitem[{Asadi et~al.(2015)Asadi, Zaeem, Nouranian, and Baskes}]{Asadi2015}
\bibinfo{author}{E.~Asadi}, \bibinfo{author}{M.~A. Zaeem},
  \bibinfo{author}{S.~Nouranian}, \bibinfo{author}{M.~I. Baskes},
\newblock \bibinfo{journal}{Physical Review B} \bibinfo{volume}{91}
  (\bibinfo{year}{2015}) \bibinfo{pages}{024105}.
\bibitem[{Kresse and Hafner(1993)}]{kresse1993ab}
\bibinfo{author}{G.~Kresse}, \bibinfo{author}{J.~Hafner},
\newblock \bibinfo{journal}{Physical Review B} \bibinfo{volume}{47}
  (\bibinfo{year}{1993}) \bibinfo{pages}{558}.
\bibitem[{Kresse and Joubert(1999)}]{kresse1999upp}
\bibinfo{author}{G.~Kresse}, \bibinfo{author}{D.~Joubert},
\newblock \bibinfo{journal}{Phys. Rev. B} \bibinfo{volume}{59}
  (\bibinfo{year}{1999}) \bibinfo{pages}{1758--1775}.
\bibitem[{Bl{\"{o}}chl(1994)}]{blochl1994projector}
\bibinfo{author}{P.~E. Bl{\"{o}}chl},
\newblock \bibinfo{journal}{Physical Review B} \bibinfo{volume}{50}
  (\bibinfo{year}{1994}) \bibinfo{pages}{17953--17979}.
\bibitem[{Perdew et~al.(1996)Perdew, Burke, and
  Ernzerhof}]{perdew1996generalized}
\bibinfo{author}{J.~P. Perdew}, \bibinfo{author}{K.~Burke},
  \bibinfo{author}{M.~Ernzerhof},
\newblock \bibinfo{journal}{Physical Review Letters} \bibinfo{volume}{77}
  (\bibinfo{year}{1996}) \bibinfo{pages}{3865--3868}.
\bibitem[{Sutton(1995)}]{sutton1995interfaces}
\bibinfo{author}{A.~P. Sutton},
\newblock \bibinfo{journal}{Monographs on the Physice and Chemistry of
  Materials}  (\bibinfo{year}{1995}) \bibinfo{pages}{414--423}.
\bibitem[{Yuasa and Mabuchi(2010)}]{yuasa2010a}
\bibinfo{author}{M.~Yuasa}, \bibinfo{author}{M.~Mabuchi},
\newblock \bibinfo{journal}{Journal of Physics Condensed Matter}
  \bibinfo{volume}{22} (\bibinfo{year}{2010}).
\bibitem[{Suzudo et~al.(2020)Suzudo, Ebihara, and Tsuru}]{Suzudo2020}
\bibinfo{author}{T.~Suzudo}, \bibinfo{author}{K.~Ebihara},
  \bibinfo{author}{T.~Tsuru},
\newblock \bibinfo{journal}{AIP Advances} \bibinfo{volume}{10}
  (\bibinfo{year}{2020}) \bibinfo{pages}{115209}.
\bibitem[{Momma and Izumi(2011)}]{momma2011vesta}
\bibinfo{author}{K.~Momma}, \bibinfo{author}{F.~Izumi},
\newblock \bibinfo{journal}{Journal of Applied Crystallography}
  \bibinfo{volume}{44} (\bibinfo{year}{2011}) \bibinfo{pages}{1272--1276}.

\end{thebibliography}

\end{document}


\begin{frontmatter}

\title{Supplemental material of ``Artificial neural network molecular mechanics of iron grain boundaries''}

\author[1]{Yoshinori Shiihara\corref{cor1}}
\ead{shiihara@toyota-ti.ac.jp}
\author[1]{Ryosuke Kanazawa}
\author[2]{Daisuke Matsunaka}
\author[3]{Ivan Lobzenko}
\author[3]{Tomohito Tsuru}
\author[4]{Masanori Kohyama}
\author[5]{Hideki Mori}

\address[1]{Graduate School of Engineering, Toyota Technological Institute, Nagoya, Aichi 468-8511, Japan}
\address[2]{Department of Mechanical Systems Engineering, Shinshu University, 4-17-1 Wakasato, Nagano, Nagano 380-8553, Japan}
\address[3]{Nuclear Engineering Research Center, Japan Atomic Energy Agency, Tokai-mura, Ibaraki 319-1195, Japan}
\address[4]{Research Institute of Electrochemical Energy, Department of Energy and Environment, National Institute ofAdvanced Industrial Science
and Technology, Ikeda, Osaka 563-8577, Japan}
\address[5]{Department of Mechanical Engineering, College ofIndustrial Technology, 1-27-1 Nishikoya, Amagasaki, Hyogo 661-0047, Japan}

\cortext[cor1]{Corresponding author}

\end{frontmatter}

\linenumbers

\section{Artificial neural network potential}
For the artificial neural network (ANN) potential, we used the high-dimensional neural network potential proposed by Behler, \textit{et al}\cite{Behler2007, Artrith2017}. It uses a feed-forward neural network with the environment of an atom in the input layer and the energy of that atom in the output layer. The training data is a set of pairs of atomic structures and their total energies obtained by the density functional theory (DFT). The parameters in the network are determined in such a way that the neural network reproduces the training data. To incorporate the atomic structure as a descriptor in the input layer, we used the Chebyshev polynomial proposed by Artrith, \textit{et al}\cite{Artrith2017} with some modifications. The modifications result in faster calculations and the details can be found elsewhere (H. Mori, to be submitted). The neural network for iron atoms was constructed by one of the authors for the analysis of the motion of dislocations in Fe crystals\cite{Mori2020}. As with conventional model potentials, the atomic systems considered as training data are essential. The atomic systems considered for the training data are ideal crystals (bcc, fcc, hcp, simple cubic structures) and these structures with disturbed cell sizes and atomic arrangements. The training data also involves those of (100), (110), (111), (112) surfaces, structures including vacancy, divacancy, and self-interstitial, and generalized stacking fault curves in (110) and (112). The neural network constructed with these datasets reproduced the reference DFT data well in terms of the bulk property of the ideal crystal, the structure of the system including defects, and the energy. Note that no grain-boundary data are included in the training data. For details, we refer to Ref. \cite{Mori2020}. The ANN package \textipa{\ae}net\cite{Artrith2016} was used for training, together with the L-BFGS optimization method\cite{byrd1995limited}. We also used the large-scale MD software, LAMMPS\cite{plimpton1995fast}, and the \textipa{\ae}net wrapper implemented in it. The aenet wrapper and ANN potential files for LAMMPS used in the simulations are available online (https://github.com/HidekiMori-CIT/aenet-lammps). The potential file, Fe.10sw-10sw.nn, can be found at ANN/Fe\_v03 directory.

\section{Grain boundary, surface, and separation energy calculations}
The symmetric-tilt grain boundary structures considered here are summarized in Table 1. The rotation axis is [110]. An example of the structures is shown in Fig. 1 along with the definition of the misorientation angle. Its corresponding surface structure is also shown as an example of the surface structures. We prepared the surface structures so that the boundary interfaces are exposed on the surface. 

We calculated GB and surface energies using the total energies obtained in these structures. The GB energy $\gamma_\text{GB}$ is defined as 
\begin{equation}
\gamma_{\text{GB}}=\frac{E_{\text{GB}}-N_\text{GB}E_{\text{bulk}}}{2A},
\end{equation}
where $E_\text{GB}$ and $A$ indicate the total energy of the GB structure and its GB area, respectively. $N_\text{GB}$ is the number of atoms involved in the structure. $E_{\text{bulk}}$ is the atomic energy for the bulk structure. Similarly, the surface energy $\gamma_{\mathrm{surf}}$ is given as the energy loss due to the lattice defect:
\begin{equation}
\gamma_{\mathrm{surf}}=\frac{E_{\mathrm{surf}}-N_{\mathrm{surf}}E_{\mathrm{bulk}}}{2A},
\end{equation}
where $E_\text{surf}$ and $N_\text{surf}$ are the total energy and the number of atoms of the surface structure, respectively. Using these two energies, the separation energy $W_{\mathrm{sep}}$ is computed as 
\begin{equation}
W_{\mathrm{sep}}=2\gamma_{\mathrm{surf}}-\gamma_{\mathrm{GB}},
\end{equation}
which is the work ideally required to separate two grains. 

\begin{figure}[t]
\includegraphics[clip,width=12cm]{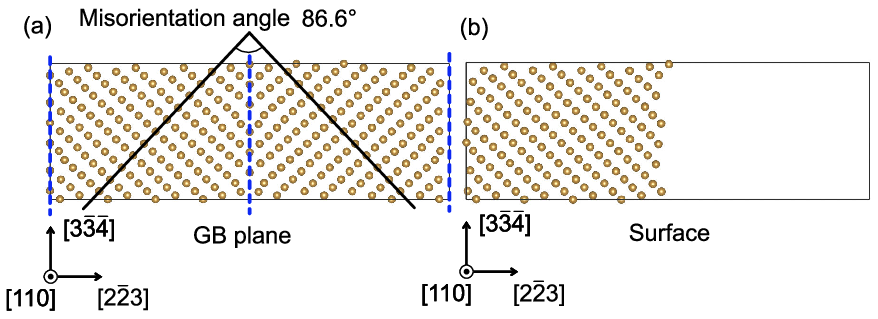}
\caption{Examples of GB and surface models: (a) Symmetric-tilt GB, $\Sigma$ 17 $(2\bar{2}3)$, structure with the definition of misorientation angle, and (b) its corresponding surface structure after atomic relaxation.}
\label{GBE_rotangle}
\end{figure}

\begin{table*}
\caption{\label{tab1} Symmetric-tilt grain boundaries considered in this study.}
\scalebox{0.9}{
\begin{tabular}{crrr}
\multicolumn{1}{c}{$\Sigma$ value} &
\multicolumn{1}{c}{GB plane} &
\multicolumn{1}{c}{Misorientation angle [$^{\circ}$]} &
\multicolumn{1}{c}{Number of atoms} \\ 
\hline      
\hline                                   
$\Sigma$ 99 &(1 $\bar{1}$ 14) &11.6&214 \\
$\Sigma$ 73 &(1 $\bar{1}$ 12) &13.5&364 \\
$\Sigma$ 51 &(1 $\bar{1}$ 10) &16.1&308 \\
$\Sigma$ 83 &(1 $\bar{1}$ 9) &17.9&556 \\
$\Sigma$ 33 &(1 $\bar{1}$ 8) &20.1&248 \\
$\Sigma$ 51 &(1 $\bar{1}$ 7) &22.9&434 \\
$\Sigma$ 19 &(1 $\bar{1}$ 6) &26.6&188 \\
$\Sigma$ 27 &(1 $\bar{1}$ 5) &31.6&324 \\
$\Sigma$ 89 &(2 $\bar{2}$ 9) &34.9&588 \\
$\Sigma$ 09 &(1 $\bar{1}$ 4) &39.0&132 \\
$\Sigma$ 57 &(2 $\bar{2}$ 7) &44.1&464 \\
$\Sigma$ 59 &(3 $\bar{3}$ 10) &46.0&348 \\
$\Sigma$ 11 &(1 $\bar{1}$ 3) &50.5&200 \\
$\Sigma$ 41 &(3 $\bar{3}$ 8) &55.9&280 \\
$\Sigma$ 33 &(2 $\bar{2}$ 5) &59.0&372 \\
$\Sigma$ 97 &(5 $\bar{5}$ 12) &61.1&424 \\
$\Sigma$ 67 &(3 $\bar{3}$ 7) &62.5&512 \\
$\Sigma$ 03 &(1 $\bar{1}$ 2) &70.6&240 \\
$\Sigma$ 81 &(4 $\bar{4}$ 7) &77.9&556 \\
$\Sigma$ 43 &(3 $\bar{3}$ 5) &80.6&420 \\
$\Sigma$ 57 &(5 $\bar{5}$ 8) &83.0&348 \\
$\Sigma$ 17 &(2 $\bar{2}$ 3) &86.6&280 \\
$\Sigma$ 99 & (7 $\bar{7}$ 10) &89.4&216 \\
$\Sigma$ 99 &(5 $\bar{5}$ 7) &90.6&616 \\
$\Sigma$ 17 &(3 $\bar{3}$ 4) &93.4&376 \\
$\Sigma$ 57 &(4 $\bar{4}$ 5) &97.1&496 \\
$\Sigma$ 43 &(5 $\bar{5}$ 6) &99.4&312 \\
$\Sigma$ 81 &(7 $\bar{7}$ 8) &102.1&436 \\
$\Sigma$ 03 &(1 $\bar{1}$ 1) &109.5&216 \\
$\Sigma$ 67 &(7 $\bar{7}$ 6) &117.6&380 \\
$\Sigma$ 97 &(6 $\bar{6}$ 5) &119.0&652 \\
$\Sigma$ 33 &(5 $\bar{5}$ 4) &121.0&272 \\
$\Sigma$ 41 &(4 $\bar{4}$ 3) &124.1&436 \\
$\Sigma$ 11 &(3 $\bar{3}$ 2) &129.5&328 \\
$\Sigma$ 59 &(5 $\bar{5}$ 3) &134.0&464 \\
$\Sigma$ 57 &(7 $\bar{7}$ 4) &136.0&324 \\
$\Sigma$ 09 &(2 $\bar{2}$ 1) &141.1&188 \\
$\Sigma$ 89 &(9 $\bar{9}$ 4) &145.1&416 \\
$\Sigma$ 27 &(5 $\bar{5}$ 2) &148.4&232 \\
$\Sigma$ 19 &(3 $\bar{3}$ 1) &153.5&280 \\
$\Sigma$ 51 &(7 $\bar{7}$ 2) &157.2&324 \\
$\Sigma$ 33 &(4 $\bar{4}$ 1) &160.0&372 \\
$\Sigma$ 83 &(9 $\bar{9}$ 2) &162.1&210 \\
$\Sigma$ 51 &(5 $\bar{5}$ 1) &163.9&464 \\
$\Sigma$ 73 &(6 $\bar{6}$ 1) &166.6&556 \\
$\Sigma$ 99 &(7 $\bar{7}$ 1) &168.5&648 \\
\hline
\end{tabular}
}
\end{table*}